\documentstyle[12pt]{article}
\begin{document}

\begin{titlepage}
\hfill CERN-TH.6844/93\\
\begin{centering}
\vfill

{\LARGE \bf Computing Masses from \\}
\vspace{0.2cm}
{\LARGE \bf
   Effective Transfer Matrices}

\vspace{2cm}
   {\bf M.\ Hasenbusch$^1$, K.\ Pinn$^2$
        and K.\ Rummukainen$^1$ } \\[6mm]

\vspace{1cm}
{\em $^1$Theory Division, CERN,\\ CH-1211 Geneva 23, Switzerland}

\vspace{0.3cm}
{\em $^2$
\,Institut f\"ur Theoretische Physik I, Universit\"at M\"unster,
      \\
      Wilhelm-Klemm-Str.\ 9, W-4400 M\"unster, Germany
      \\}

\vspace{2cm}
{\bf Abstract} \\
\end{centering}
\vspace{0.2cm}
We study the use of effective transfer matrices for the numerical
computation of masses (or correlation lengths) in lattice spin models.
The effective transfer matrix has a strongly reduced number of
components. Its definition is motivated by a renormalization group
transformation of the full model onto a 1-dimensional spin model. The
matrix elements of the effective transfer matrix can be determined by
Monte Carlo simulation. We show that the mass gap can be recovered
exactly from the spectrum of the effective transfer matrix. As a first
step towards application we performed a Monte Carlo study for the
2-dimensional Ising model. For the simulations in the broken phase we
employed a  multimagnetical demon algorithm.  The results for the
tunnelling correlation length are particularly encouraging.

\vspace{0.3cm}\noindent

\vfill \vfill
\noindent
CERN-TH.6844/93\\
March 1993
\end{titlepage}

%
\newcommand{\nc}{\newcommand}
\nc{\be}{\begin{equation}}
\nc{\ee}{\end{equation}}
\nc{\bea}{\begin{eqnarray}}
\nc{\eea}{\end{eqnarray}}
\nc{\rbo}{\raisebox}
\nc{\Nt}{N_{\tau}}
\nc{\Ns}{N_{\sigma}}
\nc{\vx}{\vec x}
\nc{\cH}{{\cal H}}
\nc{\Tr}{{\rm Tr}}
\nc{\Heff}{{\cal H}_{{\rm eff}}}
\nc{\Teff}{T_{{\rm eff}}}
\nc{\Teffl}{T_{{\rm eff}}^{(l)}}
\nc{\xna}{\xi_{0,a}}
\nc{\xes}{\xi_{1,s}} \nc{\xea}{\xi_{1,a}}
\nc{\xzs}{\xi_{2,s}} \nc{\xza}{\xi_{2,a}}
\nc{\RR} {\rangle \! \rangle}
\nc{\LL} {\langle \! \langle}
\nc{\rmi}[1]{{\mbox{\small #1}}}
\nc{\pcan}{P_\rmi{can}}
\nc{\pmm}{P_\rmi{mm}}
\nc{\ndem}{n_D}
\nc{\eq}{eq.~}
\nc{\nr}[1]{(\ref{#1})}
\nc{\ul}{\underline}
\nc{\cM}{{\cal M}}
\nc{\mc}{\multicolumn}
\thispagestyle{empty}
\newpage

\section{Introduction}
\label{intro}
The computation of the mass spectrum is one of the major goals of modern
quantum field theory. In the Euclidean path integral formulation on the
lattice \cite{Euclid}, the mass spectrum can be recovered from the
large-distance behaviour of suitably chosen correlation functions.  The
correlation functions can (in principle) be computed with the Monte
Carlo method \cite{Binder}.

The bridge between the path integral and the Hamiltonian formulation is in
the transfer matrix \cite{Euclid}. The mass spectrum can
be directly read off from the eigenvalues of the transfer matrix. The
task is therefore to diagonalize the transfer matrix. An exact solution
of the full problem is possible only in a restricted class of models.
The most prominent example for a model that can be solved via the
transfer matrix approach is the 2-dimensional Ising model
\cite{Onsager}. A direct numerical diagonalization of the transfer
matrix is restricted to systems with very small spatial extension
\cite{Numtrans}.

As the origin of the problem
with the transfer matrix approach is in the huge size of the matrix to
diagonalize, it is natural to reduce the number of degrees of freedom by
some `coarse-graining procedure'.  Inspired by block spin
renormalization group ideas \cite{Wilson}, we define an `effective
transfer matrix'.\footnote{In the literature, the notion of an
`effective transfer matrix' is used in a variety of different contexts.
The reader might be interested to compare our notion with that
developed, e.g. in \cite{Cardy}, \cite{Privman} and \cite{Borgs}.
Furthermore, for one of our definitions of the effective transfer matrix,
our approach is very similar to a method that has been discussed
in the context of glueball mass calculations \cite{Similar,Kronfeld}.
See also Appendix C.}
This effective transfer matrix has a drastically reduced size and can be
diagonalized with standard numerical procedures.

In section \ref{motivation} we motivate the concept of the effective
transfer matrix, starting from a block spin renormalization group
point of view. We arrive at an intuitive derivation of three different
`rules' for a definition of an effective transfer matrix.  In section
\ref{recover} we show (for one of the rules) that the mass of the
first excited state can be exactly recovered from the effective
transfer matrix. To learn more about the properties of the effective
transfer matrix, we then study the 2-dimensional Ising model on small
lattices.  Here we can compute the spectrum of the effective transfer
matrix without any approximation and compare it with the exactly known
spectrum of the full transfer matrix. So we can study to what extent
the low-lying spectrum of the full transfer matrix can be recovered
from the effective transfer matrix. In the final section we present
results for the low-lying spectrum of the 2-dimensional Ising model on
lattices up to $64 \times 2048$ as obtained from effective transfer
matrices computed by Monte Carlo simulations.  We discuss results from
simulations at the critical point and in the broken phase of the
model.  The multimagnetical demon algorithm we employed for the
simulations in the broken phase provided us with high statistics for
tunnelling events. The success is based on an increased probability
for configurations with low magnetization and a high performance due
to a very efficient implementation.  The longest tunnelling
correlation length we measured was $43500\pm 1600$ on a $64\times
128$ lattice, consistent with the exact value $44014.4\ldots$.

\section{Effective 1-Dimensional Model}
\label{motivation}
We consider a $(d+1)$-dimensional Euclidean quantum field theory on a
cubic lattice $\Lambda$ of size $\Ns^d \times \Nt$.  Here, $\Ns$ denotes
the spatial extension of the lattice, and $\Nt$ is the extension in the time
direction.  Let us label the sites of the lattice by $(t,\vx)$, where
$t$ runs from $1$ to $\Nt$, and the spatial coordinates $x_i$ cover the
range $1 \dots \Ns$.  In the Euclidean path integral formulation one
integrates over stochastic variables $\phi_{t,\vx}$ that are attached to
the sites of $\Lambda$.  The model is defined by the
partition function
\be
 Z = \sum_{\phi} \exp\bigl( - \cH(\phi) \bigr) \, .
\ee
Let us denote the configurations of $\phi$ on `time slices' $t$
by $\phi_t$. Assume that $\cH(\phi)$ takes the following form:
\be
 \cH(\phi) = \sum_t \bigl( K(\phi_t,\phi_{t+1}) + V(\phi_t) \bigr) \, .
\ee
Then for periodic boundary conditions in the $t$-direction
the partition function may be written as
\be
 Z = \Tr \bigl( T^{\Nt} \bigr) \, .
\ee
The transfer matrix $T$ can be chosen to be symmetric:
\be
 T(\varphi,\varphi') =
 \exp \bigl( - K( \varphi,\varphi' )
  - \frac12 [ V(\varphi) + V(\varphi') ] \bigr) \,
\ee
$\varphi$ and $\varphi'$ are again time slice configurations, i.e.\
configurations on $d$-dimensional sub-lattices of size $\Ns^d$.
We shall also use a `bra' and `ket' notation, e.g.
\be
 T(\varphi,\varphi') = \langle \varphi \vert T \vert \varphi' \rangle \, .
\ee
Let $\lambda_i$, $i=0,1,\dots$ denote the eigenvalues of $T$, such
that $\lambda_0 > \lambda_1 \geq \lambda_2 \geq \dots $.
The corresponding eigenvectors shall be denoted by
$\vert i \rangle$. For eigenstates with zero momentum, masses
are defined as
\be
 m_i = - \ln \frac{\lambda_i}{\lambda_0} \equiv \frac{1}{\xi_i}
\, .
\ee
The $\xi_i$ denote the corresponding correlation lengths.
We shall now consider block spin transformations that transform
the statistical system from a $(d+1)$-dimensional one to a
$1$-dimensional one. Block spins are defined as averages over
time slices:
\be
 \Phi_t \equiv \Ns^{-d} \sum_{\vx} \phi_{t,\vx}  \, .
\ee
For integer $l \geq 1$ we define an `effective Hamiltonian'
\be
 \exp\bigl(-\Heff^{(l)}(\Phi)\bigr)=
 \sum_{\phi} \exp\bigl( -\cH(\phi) \bigr)
 \prod_{t \in G_l} \, \delta
 \bigl( \Phi_t - \Ns^{-d} \sum_{\vx} \phi_{t,\vx}  \bigr) \, ,
\ee
with  $ G_l = \{ 1,l+1,2l+1,\dots,\Nt-l+1 \}$, and
$Z$ can be rewritten as
\be
 Z = \left( \prod_{t \in G_l} \sum_{\Phi_t} \right) \,
 \exp \left( -\Heff^{(l)}(\Phi) \right) \, .
\ee
Let us now relabel the sites in $G_l$ with integers
that run from $1$ to $\Nt'=\Nt/l$.
Let us  furthermore assume that (after the relabelling) for
some $l$ with good precision $ \Heff^{(l)}(\Phi) $ couples only
nearest neighbours, i.e.\ is of the form
\be
 \Heff^{(l)}(\Phi) =
 \sum_t \left( K_{{\rm eff}}^{(l)}(\Phi_t,\Phi_{t+1}) +
 V_{{\rm eff}}^{(l)}(\Phi_t) \right) \, .
\ee
Note that, in general, the effective Hamiltonian will contain
interaction terms of arbitrary range. However, one can generally
assume that these terms decay exponentially with the distance
and might be neglected for a first qualitative analysis.

Assuming a nearest neighbour interaction effective Hamiltonian, we can
again rewrite the partition function in terms of a transfer matrix
\be
 Z = \Tr \left( \lbrack \Teffl \rbrack^{\Nt'} \right) \, ,
\ee
 and
\be
 \Teffl(\Phi,\Phi') =
 \exp \left( - K_{{\rm eff}}^{(l)}( \Phi,\Phi' )
 - \frac12 [ V_{{\rm eff}}^{(l)}(\Phi)
  + V_{{\rm eff}}^{(l)}(\Phi') ] \right) \, .
\ee
What have we gained?  The number of degrees of
freedom has been considerably reduced.  Take for example the Ising
model.  Here the original transfer matrix is $2^{\Ns^d}$ by $2^{\Ns^d}$.
If we choose as block spin the time slice magnetization then $\Teffl$ is
a matrix with $\Ns^d + 1$ by $\Ns^d +1 $ components.  So there is a
drastic simplification in the eigenvalue problem.  The crucial
question is, of course, to which extent the spectrum of $\Teffl$ reflects
properties of the original system.  This question will be studied in
detail in the next section.

Let us first discuss how one can express the elements of $\Teffl$ as
expectation values in the statistical system.  This is necessary if we
want to compute them, e.g. by Monte Carlo simulation.  To simplify
the discussion, we shall assume in the following that the $\Phi_t$
take discrete values only.  Let us denote the discrete values of
$\Phi_t$ by $M,N,\dots$.  They correspond to `improper states' $\vert
M \RR$,$\vert N \RR$, etc. We shall use throughout `double brackets'
to denote states in the space that $\Teffl$ acts on.  We now introduce
the operator $\delta_{\Phi_t,M}$ that takes the value $1$ if $\Phi_t =
M $, and $\delta_{\Phi_t,M} =0 $ elsewhere.  Assuming periodic boundary
conditions in the $t$-direction, the correlator of two such operators at
distance one can be written as follows:
\be\label{bubb}
 <\delta_{\Phi_t,M} \delta_{\Phi_{t+1},N}> =
 Z^{-1}  \LL M|\Teffl|N \RR \LL N| \lbrack \Teffl
 \rbrack^{N_{\tau}'-1}|M \RR \, .
\ee
In the following we shall sometimes omit the extra factor $Z^{-1}$,
because it leads only to an irrelevant shift of the ground-state
energy. Masses and correlation lengths are unaffected.  Equation
(\ref{bubb}) has to be resolved with respect to the effective transfer
matrix elements.  This is particularly easy for the case
$N_{\tau}'=2$,
\be
 \LL M|\Teffl|N \RR =
 \sqrt{<\delta_{\Phi_1,M} \delta_{\Phi_{2},N}>  } \, .
\ee
We say that the effective transfer matrix is defined according to the
{\em symmetric periodic lattice rule}.  For larger $N_{\tau}'$ ({\em
asymmetric periodic lattice rule}), the solution for $\Teffl$ can be
found by iteration (cf.\ Appendix A).  The situation becomes
simple again in the limit $N_{\tau}' \rightarrow \infty$,
\be
 <\delta_{\Phi_t,M} \delta_{\Phi_{t+1},N}> =
 \LL 0|N \RR \LL N|\Teffl|M \RR\LL M|0 \RR \, .
\ee
Using the fact that $<\delta_{\Phi_t,M}> = \LL 0|M \RR\LL M|0 \RR$, one finds
\be\label{thethe}
 \LL M  \vert \Teffl \vert N \RR =
 \frac {<\delta_{\Phi_t,M} \delta_{\Phi_{t+1},N}>}
 {\sqrt{<\delta_{\Phi_t,M}><\delta_{\Phi_t,N}>}} \, .
\ee
We say that $\Teffl$ is defined here via the {\em $\Nt=\infty$ rule}.
\section{How the Mass Gap is Recovered}
\label{recover}
In this section we want to discuss how the effective transfer matrix
$\Teffl$ is related to the transfer matrix of the basic system, $T$, and
how much of the spectrum of this matrix can be recovered from the
spectrum of the effective transfer matrix. We will concentrate on the
discussion of the $\Nt = \infty$ rule.

We define `magnetization pieces of the ground-state' by
\be\label{pieces}
 \vert M \rangle = {\cal N}_M^{-1} \sum_{\varphi}
 \delta \left( \bar \varphi = M \right) \,
 \vert \varphi \rangle \langle \varphi \vert 0 \rangle  \, .
\ee
Here, $\vert \varphi \rangle$ are time slice configuration
states, and
\be
 \bar \varphi = \Ns^{-d} \sum_{\vx} \varphi_{\vx} \, .
\ee
The constants ${\cal N}_M$ are chosen such that
$\langle M \vert M \rangle = 1$.
One can convince oneself that
$ \langle N \vert M \rangle = \delta_{N,M}$.
However, the $\vert M \rangle $'s do not, of course,
span the complete Hilbert space. Let us denote
by ${\cal P}$
the projector onto the subspace spanned by the
$\vert M \rangle$'s ,
\be
 {\cal P}= \sum_{M} \vert M \rangle \langle M \vert  \, .
\ee
The effective transfer matrix for the
$\Nt = \infty$ rule is given by

\be
 \LL M \vert \Teffl \vert N \RR =  \langle M \vert T^l \vert N \rangle \, .
\ee
Let us denote the eigenstates of $\Teffl$ by $\vert i \RR$,
and the corresponding eigenvalues by $\Lambda_i^{(l)}$.
We now claim that
 $\Lambda_0^{(l)}= \lambda_0^l$. The proof is as follows:
\bea
 \LL M \vert \Teffl \vert 0 \RR &=&
 \sum_N  \LL M \vert \Teffl \vert N \RR \LL N \vert 0 \RR \nonumber \\
 &=& \sum_N  \langle  M \vert T^l \vert N \rangle
 \langle N \vert 0 \rangle \nonumber \\
&=& \langle M \vert T^l {\cal P} \vert 0 \rangle \nonumber \\
&=& \lambda_0^l \,  \LL M \vert 0 \RR \, .
\eea
Here we have used the fact that ${\cal P} \vert 0 \rangle = \vert 0 \rangle $.

What about the first excited state?  Let us try the following wave
function:
\be
 \LL M \vert 1 ' \RR = c_1^{-1/2} \langle M \vert 1
\rangle \, .
 \ee
  The constant $c_1$ is chosen such that this state is
normalized to 1:
 \be\label{c1}
  c_{1} = \sum_M \langle M|1 \rangle^2 =
\langle 1 \vert {\cal P} \vert 1 \rangle \, . \ee
 The action of the
effective transfer matrix on this state is easily found to be
\be\label{ffff}
 \LL M \vert \Teffl \vert 1' \RR = c_1^{-1/2} \langle M
\vert T {\cal P} \vert 1 \rangle \, .
 \ee
 Now consider the first excited
state $|1 \rangle $ of the original transfer matrix $T$ with eigenvalue
$\lambda_1$ and rewrite it as follows:
 \be |1 \rangle = c_{1}^{-1}
{\cal P} \vert 1 \rangle + \sum_{i > 1} f_{1,i} | i \rangle \, .
 \ee
That only states with $i \!>\! 1$ contribute in the sum is due to the
fact that $ \langle 0 \vert {\cal P} \vert 0 \rangle = 1$, and $ \langle
0 \vert {\cal P} \vert 1 \rangle = 0$.  Solving this equation for ${\cal
P} \vert 1 \rangle$ and inserting this in eq. (\ref{ffff}), one
obtains
 \be
  \LL M \vert \Teff^{(l)} \vert 1' \RR = c_1 \lambda_1^l \left(
\LL M \vert 1' \RR - \sum_{i > 1} \frac{f_{1,i}}{c_1^{1/2}} \left(
\frac{ \lambda_i}{\lambda_1}\right)^l \langle M \vert i \rangle \right)
\, .
 \ee
For large $l$ one therefore has
 \be
  \Teff^{(l)} |1' \RR = {c_1}
\lambda_1^l |1' \RR + O \left( (\frac{\lambda_2}{\lambda_1})^l \right)
\, .
\ee
{F}rom the spectrum of the effective transfer matrix, we obtain
estimates for the mass of the first excited state
\be\label{xieffl}
 m_1^{(l)} \equiv -
\frac1l \ln \left( \frac{ \Lambda_1^{(l)}} {\Lambda_0^{(l)} } \right)
\equiv \frac{1}{\xi_1^{(l)}} \, .
\ee
It is easy to see that $m_1^{(l)}$
behaves like
\be\label{form}
 m_1^{(l)} = m_1 - \frac{\ln c_1}{l} +
\mbox{\ exponentially small corrections} \, .
\ee
 One can get rid of the
$1/l$ corrections by combining pairs of results with different $l$'s.
For the correlation length of the first excited state one gets
\be\label{combine}
 \xi_1 = \frac{ l_1 \xi_1^{(l_1)} - l_2 \xi_1^{(l_2)}
}{l_1 - l_2} + \mbox{\ exponentially small corrections} \, .
\ee
What about the higher states?
Let us first see what happens with the second
excitation.  Proceeding na\"\i vely, we would start with
\be \LL M \vert 2'
\RR = c_2^{-1/2} \langle M \vert 2 \rangle \, ,
\ee
and go along the
same lines as in the proof above.  We would then find that everything
would work just the same as for the first excited state, provided that
$\langle 1 \vert {\cal P} \vert 2 \rangle$ vanishes.  (This can happen
for symmetry reasons, see the discussion of the Ising model below.)
Generally, when for higher states the overlaps $\langle i \vert {\cal P}
\vert j \rangle$ are very small for $i \!  < \! j$, then the state
$\vert i' \RR$ allows the reconstruction of $\lambda_i$ with good
precision.

\section{Exact Results on Small Lattices}
\label{exact}
As a first study of the effective transfer matrices defined above, we
made numerical calculations for the 2-dimensional Ising model on
lattices with small spatial extension.  These calculations are exact in
the sense that only standard numerical techniques were used.  No Monte
Carlo simulations were involved.

The model is defined through its partition function
\be
 Z = \sum_{\sigma_{x} =\pm 1}
 \exp \left( \beta \sum_{ \langle x,y \rangle} \sigma_{x} \sigma_{y} \right) \,
{}.
\ee
Here we have denoted lattice sites by $ x = (t,i)$, and
$ \langle x y \rangle $ denotes a nearest neighbour pair.
The (symmetrically chosen) transfer matrix of this model is given by
\be\label{trama}
 T(\phi_{t} ,\phi_{t+1} ) =
 \exp \left(
    \frac{\beta}2 \sum_{i=1}^{\Ns} \sigma_{t,i} \sigma_{t,i+1}
  + \beta \sum_{i=1}^{\Ns} \sigma_{t,i} \sigma_{t+1,i}
  + \frac{\beta}2 \sum_{i=1}^{\Ns}
             \sigma_{t+1,i} \sigma_{t+1,i+1}  \right) \, ;
\ee
$\phi$ here denotes time slice configurations of the Ising spins.
We assume periodic boundary conditions in the space direction,
i.e.\ the site with $i=\Ns+1$ is identified with
the site $i=1$.

The 2-dimensional Ising model was first solved by Onsager in 1944
\cite{Onsager}. The eigenvalues of the transfer matrix can be read off
easily from \cite{Kaufman}.  See also the reviews in ref. \cite{Reviews}.

As a consequence of the fact that the transfer matrix commutes with the
spatial translation operator, the eigenstates can be chosen as
simultaneous eigenstates of energy and momentum.  They can further be
classified in states that are symmetric/antisymmetric with respect to
reversal of all spins.  We shall denote the corresponding masses or
correlation lengths with a subscript $s$ or $a$, respectively.

In the infinite volume limit, i.e.\ when $\Ns \rightarrow \infty$, the
spectrum is twofold degenerate in the broken phase ($\beta > \beta_c$).
For finite $\Ns$ the degeneracy in the
low-temperature regime is lifted and level splitting occurs because
 of  tunnelling.

We will consider the time slice magnetization as effective spin in the
following. In this case only the zero-momentum eigenvalues
can be recovered from the effective transfer matrix.
\footnote{Note that effective states that couple to nonzero
momentum could also be employed}

We shall give results for the $\Nt=\infty$ rule and for the symmetric
periodic lattice rule.  We have results for $N_{\sigma} \leq 8$.  The
basis for our computations is an accurate determination of the
eigenvalues and eigenfunctions of the transfer matrix defined in eq.\
(\ref{trama}).  In principle the eigenfunctions of $T$ are given in the
work by Kaufmann \cite{Kaufman}.  However, the expressions seem too
complicated to be of practical use for our goals, so we decided to use
the computer.

Let us first discuss the case $\Nt = \infty$.
We numerically determined the eigenstates $\vert i \rangle$ and
the `magnetization pieces' $\vert M \rangle$ defined in
eq.\ (\ref{pieces}). Together with the eigenvalues of $T$, they
enter the matrix elements of $\Teffl$ as follows:
\be
 \LL M \vert \Teffl \vert N \RR  =
\sum_{i} \lambda_{i}^l \langle M| i \rangle  \langle i | N \rangle \, .
\ee
The $\vert M \rangle$'s are
zero momentum states.  Hence we only needed to
calculate the translational invariant eigenstates of $T$.

In the case of the periodic system with $\Nt = 2 l$
(symmetric periodic lattice rule) one finds
\be
\LL M \vert \Teffl \vert N \RR  =
  \sqrt{\sum_{i,j} \lambda_{i}^l \lambda_{j}^l
  \langle i | M \rangle \langle M | j \rangle
  \langle j | N \rangle \langle N | i \rangle }  \, .
\ee
For both rules we computed the spectrum of $\Teffl$ for various

$l$ and determined the $\xi^{(l)}$ defined in
eq.\ (\ref{xieffl}) for the lowest-lying
states. Our results for the $\Nt=\infty$ rule are listed in
tables \ref{ns4infty} and \ref{ns8infty}. The results
for the symmetric periodic lattice rule are quoted in tables
\ref{ns4periodic} and \ref{ns8periodic}. We always quote
the results for $l=1$, 2, 4 and 8. In addition, we give
an estimate for the `true' $\xi$ obtained by combining
$\xi^{(4)}$ and $\xi^{(8)}$ according to eq.~(\ref{combine}).
The exact results are always quoted with an `e' in the
second column.

Let us first look at the results for the $\Nt=\infty$ rule.
As we proved in section \ref{recover}, the
$\xna^{(l)}$ should converge towards the exact limit $\xna$.
This can indeed be observed.
The second largest correlation length $\xes$ also nicely converges.
This is a consequence of the fact that the corresponding state is
symmetric in the magnetization while the first
excited state is antisymmetric.
Therefore the matrix element $\langle 1s \vert {\cal P} \vert 0a \rangle$
vanishes, and the proof that the correct correlation length can be
recovered goes the same way as for the first excited state.
The convergence for $\xea$ is also quite good. This is
anticipated from the discussion at the end of section
\ref{recover}, since $c_{0a}$
is very close to 1.
Following the discussion of section \ref{recover}, the constant
$c_i = \sum_M \langle M \vert i \rangle^2$ generalizing
eq. (\ref{c1}) should indicate the goodness of the effective
eigenstate corresponding to $\vert i \rangle$;
$c_i = 1$ indicates a perfect representation of the eigenfunction
$\vert i \rangle$ by the corresponding effective eigenfunction
$\vert i' \rangle$.

For $L=4$, $6$, $8$ and $10$ we calculated this quantity explicitly,
starting from the exact eigenfunctions with labels
$i=(0,a)$,
$(1,s)$ and $(1,a)$. The results for the $\beta$-range
0.35 -- 0.55 are given in fig. 1.
We make the following observations:
The overlap $c_{0a}$ is very close to 1. It becomes better for
$\beta \rightarrow 0$ and for
$\beta \rightarrow \infty$. The overlaps
$c_{1s}$ and $c_{1a}$ are worse. The deviation from 1 is in
the range of several per cent;
$c_{1s}$ and $c_{1a}$ are `best' in the critical domain.
All $c$'s become worse with increasing $\Ns$. It is an open
question whether they have a finite limit in the
$\Ns \rightarrow \infty$ limit.

The $\xzs^{(l)}$ do not converge. Here the states of the
original system seem to mix somehow to the effective states.

Let us now turn to the discussion of the results for the periodic
symmetric rule. Table \ref{ns4periodic} is  encouraging: when we combine
the $l=4$ and $l=8$ data,  all correlation lengths
are reproduced within the given numerical accuracy,
with the
exception  of $\xzs$. However, this  is only so
on very small lattices. A careful look  at table \ref{ns8periodic}
reveals that only the leading correlation lengths $\xna^{(l)}$ converge
to the exact value $\xna$. The estimates for the other $\xi$'s are not
entirely off. They might be regarded as unsystematic approximations.
However, we do not know what happens if the spatial extension of the
lattice is further increased. See also the discussion of Monte
Carlo results on larger lattices in section \ref{MCresults}.

\section{Monte Carlo Results for the 2D Ising Model}
\label{MCresults}
\subsection{The Critical Model, $\Nt=\infty$ rule}
At the critical coupling $\beta_c = 0.4406868...$
of the 2-dimensional Ising model, we performed Monte Carlo
simulations on  $\Ns \times \Nt = 16 \times 512$, $32 \times 1024$ and
$64 \times 2048$ lattices. We always measured and stored the magnetizations
of all time slices after five
Swendsen-Wang updates \cite{Swendsenwang} of the
entire lattice.

We made 5000 measurements for $\Ns=16$, 3900 measurements for $\Ns=32$,
and 3000 measurements for $\Ns=64$.

For the simulations at criticality we used the $\Nt=\infty$ rule to get
estimates for the effective transfer matrix elements. `Infinitely long'
lattice here means, of course, that $\Nt$ is much larger than the
largest correlation length involved, so that the influence of a finite
lattice extension in the $t$-direction can be neglected.
In all three cases, $\Nt / \xi_{0,a}$
is about 25. This is certainly on the safe side.

The eigenvalues and eigenstates of the effective transfer matrix were
then calculated using standard numerical procedures. {F}rom the
eigenvalues of the effective transfer matrix we determined the
$\xi^{(l)}$ for various distances $l$.

In order to estimate the statistical errors of our results we divided
the whole sample in  5, 10, 20 and 40 bins
and computed the effective
transfer matrix and its eigenvalues separately within each of the bins.
The statistical errors quoted in the tables were obtained from the mean
square fluctuations over the outcomes of the various bins. We considered
the error estimate as reliable when it was approximately independent of
the number of bins.

This binning analysis was also used to check for a bias due to a too
small statistics. We averaged the results from the bins and compared the
outcome of various bin sizes. Only those results can be trusted where
there is an agreement of the results over the various bin sizes.
The quantities for which we obtained stable estimates are quoted in
tables \ref{mcl16}, \ref{mcl32} and \ref{mcl64}.

With increasing $l$, at least the largest two correlation
lengths $\xi_{0,a}$ and $\xi_{1,s}$ converge towards
the corresponding exact value.
Similar to the exact results on small lattices we find the
estimates for $\xea$ converging within the statistical accuracy
towards the exact results.

The tables also contain our final estimates for the correlation
lengths obtained by combining
pairs of effective correlation lengths with different
$l$ according to eq.\ (\ref{combine}).
The convergence and stability of these combined estimates
shows that the $l$~dependence is indeed of the form of eq.\ (\ref{form}).

In fig. 2 we present plots of the effective wave functions
(eigenfunctions of $\Teffl$) for $\Ns = 64$ and for
$l=5,10$ and $20$.
One can observe that the effective wave functions obtained from
different $l$ are almost identical. For $l=20$ the effective
 wave function for $ \vert 1a \rangle $ becomes noisy because
  of insufficient statistics.

In order to compare the effective transfer matrix approach with
a `traditional' determination of masses, we present here
the results of a conventional evaluation of our Monte Carlo data
for the 2-dimensional Ising model at criticality.
As an example we choose $\Nt=64$ and $\Nt = 2048$.
We define
\bea
 G_{0,a}(t) &=& < \phi_0 \phi_t> \, , \nonumber \\
 G_{1,s}(t) &=& < \phi_0^2 \phi_t^2> - < \phi_0^2 >< \phi_t^2> \, .
\eea
These correlation functions behave like
$ G_i(t) = g_i \, \exp( - t / \xi_i ) + \dots  \, .$
We define `effective correlation lengths' as
\bea
\xi^{{\rm eff},t}_i =
1 / \left( \ln(G_i(t))-\ln(G_i(t+1)) \right) \, .
\eea

With increasing $t$ these quantities should converge to the true
correlation lengths. Table \ref{conv} shows our results for the
$\xi^{{\rm eff},t}_i$, for $i=(0,a)$ and $i=(1,s)$. We used exactly the
same Monte Carlo data as for the effective transfer matrix. The
comparison of the results obtained with the two different methods shows
that the transfer matrix results have statistical errors that are
roughly a factor of two smaller than the `conventional' ones.

We evaluated the same Monte Carlo data with a third technique: Based on
the results for the eigenfunctions of the effective transfer matrix
$\Teffl$, we determined observables that are expected to have improved
overlap with the eigenstates $\vert 0,a \rangle$ and $\vert 1,s
\rangle$, respectively. For details see Appendix C. With the improved
observables, we again computed the `effective correlation lengths'
$\xi^{{\rm eff},t}_i$, for $i=(0,a)$ and $i=(1,s)$. The comparison of
these quantities, `standard' and `improved' is shown in figs. 3 and 4.
At least for $\xna$, the improvement is striking.

\subsection{Monte Carlo Results for the Broken Phase}

We also did simulations in the broken phase, at $\beta = 0.47$.
This value is low enough so that the tunnelling correlation length
becomes very large even with modest $N_\sigma$, but is close enough to
$\beta_c$ so that the bulk correlation length ($4.349\ldots$
when $\Ns = \infty$) is still substantially larger than one.

In order to fight the {\em supercritical slowing down} due to
exponentially suppressed tunnelling rates, we employed a
multimagnetical algorithm. Thanks to multispin coding and the usage of
demons to implement the nonlocal magnetization-dependent interaction
terms in the multimagnetical ensemble, the algorithm performed very
well on a CRAY X-MP. For more details, see Appendix B.  We performed
simulations for lattices with spatial extensions $\Ns= 16$, $32$ and
$64$. For an overview of the runs and the statistics see table
\ref{statbro}.

For the analysis in the broken phase, we used the {\em asymmetric
periodic lattice rule} throughout. Our results are quoted in
tables \ref{bro1}, \ref{bro2} and \ref{bro3}.
We performed the simulations with $\Nt = \Ns/2$, $\Ns$ and $2\Ns$.
Generally, we observe two trends: the larger $\Nt$ is, or
the closer one is to the symmetric periodic lattice rule, the
smaller the systematic error in the tunnelling correlation length
measurements becomes.
When $\Nt=2\Ns$, we find an impressive reproduction of the tunnelling
correlation length that gets as big as 44014 on the $\Ns=64$ lattice.
Note that the tables also include results for the symmetric periodic
lattice rule: the last row in each block of the table quotes the
results for $l=\Nt/2$.  We also present the tunnelling correlation
length data for the $\Ns=64$ lattices in fig. 5.

We also tried to reproduce the large tunnelling correlation length on
the $\Ns = 64$ lattice using the standard technique of fitting the
correlation functions of time slice magnetizations.  However, this did
not lead to any sensible result. To us this seems to be a situation
where our effective transfer matrix technique is completely superior.
We therefore consider our method well suited for the study of the
interface tension \cite{Interface}.

The other correlation lengths besides the tunnelling length are only
approximately correct. It seems that they do not converge. Here
certainly further investigation is needed.

The effective wave functions for $\beta=0.47$ and $\Ns = 64$ are
plotted in fig. 6, for $l=8,16$, and $32$.
That the tunnelling length is so large corresponds to the fact
that the square of the ground-state
wave function and the square of the first excited state differ only in
a small neighbourhood of zero magnetization.

\section{Summary and Conclusion}

We have studied the use of effective transfer matrices for the computation
of masses of a Euclidean quantum field theory, or, equivalently,
the correlation lengths of a statistical mechanical model.  Many
questions are left open and deserve further theoretical and numerical
investigation.  The theoretical study of the other rules besides
the $\Nt = \infty$ rule is not yet complete.  Furthermore, one could
consider the usage of blocks with finite extension in the time direction.  In
principle, one can employ any sort of `effective spin' (not just the
magnetization or its absolute value).  Application of the method to
systems with continuous degrees of freedom would also be interesting.

A report on a study of the 3-dimensional Ising model will be published
elsewhere \cite{3dtocome}.

\section*{Acknowledgement}
We would like to thank Christian Wieczerkowski for
the suggestion to identify block spins with
projectors. K.P. would also like to thank Christian
for helpful discussions. M.H. would like to thank
U. Wolff, M. M\"uller-Preussker and R. Ben-Av for discussions.
\section*{Appendix A: Matrix Equation for Asymmetric Rule}
The problem of solving the equation
\be
 <\delta_{\Phi_t,M} \delta_{\Phi_{t+1},N}> =
  \LL M|\Teffl|N \RR \LL N|
  \lbrack \Teffl \rbrack^{N_{\tau}'-1}|M \RR
\ee
with respect to the components of the effective transfer matrix (given
the left-hand side, e.g.\ as an outcome of a Monte Carlo simulation),
may be stated as follows:  For a given symmetric matrix $B$ with matrix
elements $B_{ij}$, find a matrix $A$ with matrix elements $A_{ij}$ such
that the following equation holds:
\be\label{abcd}
B_{ij} = A_{ij} \, (A^n)_{ij} \, ,
\ee
where $A^n$ denotes the $n^{th}$ matrix power of $A$.  We do not know of a
closed solution of this problem, but, inspired by the famous iteration
prescription for the problem of finding the square root of a number $a$,
 $x \rightarrow \frac12 \bigl( x + \frac{a}{x} \bigr)$, we try the
following iteration prescription for the solution of eq.\ (\ref{abcd}):
\be
 A_{ij} \rightarrow \frac1{1+\zeta} \bigl( A_{ij} + \zeta
 \frac{B_{ij}}{(A^n)_{ij}} \bigr) \, ,
\ee
where $\zeta$ denotes a parameter that can be tuned in order to
optimize the convergence.  With the exception of a few special cases the
algorithm converged, although the convergence rates were sometimes very
slow.
\section*{Appendix B: Multimagnetical Demon Algorithm}

We here present a brief description of the multimagnetical demon
algorithm used for the simulations at $\beta=0.47$.

When $\beta > \beta_c$, the tunnelling rate between the ordered states
becomes exponentially smaller as the size of the system is increased.
However, in order to extract the tunnelling correlation length we need
a good statistics of the tunnelling events -- i.e.\@ the
configurations with the magnetization between the bulk expectation
values $\pm \cM_B$.  The multimagnetical method \cite{Berg91} solves
this problem by artificially enhancing the probability of these
states.  In the standard (no demons) approach, this is achieved by
modifying the probability of the spin configuration $\bar\sigma$ with
an extra weight function $G(\cM)$:
\be
\pmm(\bar\sigma) \propto e^{-\beta {\cal H}(\bar\sigma)}
        \,G(\cM_{\bar\sigma}),
\label{multimag}
\ee
where $\cM_{\bar\sigma}$ is the magnetization of the configuration
$\bar\sigma$.  The probability of the magnetization $\cM$ becomes
\be
\pmm(\cM) \propto \pcan(\cM)\,G(\cM) \propto
\left(
\sum_{\bar\sigma} e^{-\beta {\cal H}
(\bar\sigma)}\delta(\cM_{\bar\sigma}-\cM)
\right)\,G(\cM) \, .
\label{magprob}
\ee
We use a notation where the $\beta$-dependence of $G$ and $P$ is
suppressed.  Usually one aims at a constant probability distribution
between $\pm \cM_B$, implying that $G(\cM)\propto 1/\pcan(\cM)$. However,
since $\pcan$ is unknown, one has to use an approximate form instead, which
can be
obtained, for example, by scaling up the function $G$ used in the
simulations performed with smaller volumes.  This is often further
refined by performing test runs and adjusting $G(\cM)$ until satisfactory
$\pmm(\cM)$ is obtained.  For further details, see refs.
\cite{Berg91,Rummukainen92}.

The function $G(\cM)$ depends on the global magnetization, rendering the
update non-local and thus preventing straightforward vectorization and
parallelization.  In this work modified the standard multimagnetical
algorithm by utilizing {\em demons\,} to change the magnetization; this
approach enables us to use powerful multispin coding and is highly
vectorizable.  This method is a magnetic analogue of the multicanonical
demon algorithm presented in \cite{Rummukainen92}.  The actual
update becomes a two-stage process: first, the demon magnetization
$\cM_D$ is changed by coupling the demons to a ``multimagnetical heat
bath'', and second, the spin system magnetization $\cM_S$ is changed by
coupling the spins to demons and performing the spin update while
preserving the total magnetization $\cM_T = \cM_D + \cM_S$.
More precisely, we perform the simulation according
to the joint probability distribution
\be
P(\cM_S,\cM_D) \propto \pcan(\cM_S)\,\ndem(\cM_D) \,W(\cM_T)\,,
\label{multidemon}
\ee
where $\ndem(\cM_D)$ is the number of demon states with magnetization
$\cM_D$, and $W$ is a new weight function that depends only on $\cM_T$.
The canonical expectation value of an observable $\cal O$ can now
be obtained by reweighting:
\be
<{\cal O}> = \frac{\sum_i {\cal O}_i W^{-1}(\cM_{T,i}) }
        {\sum_i W^{-1}(\cM_{T,i})},
\label{expectation}
\ee
where ${\cal O}_i$ and $\cM_{T,i}$ refer to the individual measurements
of the corresponding quantities.  Summing over $\cM_D$ in
\eq\nr{multidemon} and comparing it with \eq\nr{magprob}, we note that $G$
and $W$ are related:
if $G(\cM_S)\propto\sum_{\cM_D}\ndem(\cM_D)\,W(\cM_T)$,
we obtain a similar probability distribution for $\cM_S$ in the two cases.
In this work we are using demons carrying $\pm 1$ units of
magnetization; with $N_D$ demons, the demon density of states becomes
\be
\ndem(\cM_D) = \frac{N_D!}{[(N_D + \cM_D)/2]!\,[(N_D - \cM_D)/2]!}\,.
\label{demonden}
\ee
Let us look closer at the individual update steps.

$\bullet$ In the multimagnetical demon refresh step the new $\cM_D$
is chosen with probability
\be
P(\cM_D)\propto\ndem(\cM_D)\,W(\cM_S+\cM_D).
\label{demonprob}
\ee
While one could use this probability to refresh each demon
individually, this is not very efficient ($\propto N_D$ steps).  We
used \eq\nr{demonprob} to directly choose new random $\cM_D$.  Depending
on the old value of $\cM_D$, we then either added or subtracted
magnetization
from randomly selected demons, until the right demon magnetization was
reached.  This whole process takes, on the average,
only $\propto\sqrt{N_D}$ steps, and yields a new demon magnetization
independent of the old one.  In our simulations we used
four times as many demons as spins ($N_D = 4N_S$).

$\bullet$ In order to ensure the canonical energy distribution the
spin system is connected to a heat bath.  To make the multispin coding
easier, we used a second set of demons, this time carrying energy.  Individual
demon energies vary in units of 4: $0,4,8,\ldots$.  Before each update
sweep through the spin system, every spin is connected to a magnetic
demon and an energy demon; the demons are chosen with random order to
ensure fast mixing.  A spin flip is accepted if and only if the demons
can absorb the change in energy and magnetization.  The energy demons
are periodically refreshed with a heat bath.  Note that one could also
perform a normal Metropolis or heat bath update without the energy demons;
the update is accepted/rejected with the magnetic demon.

Because the spin update is very fast, we interleaved 5 sweeps through
the lattice for one demon update. For the largest lattice
($128\times64$), an individual spin update took 16 ns, whereas the total
time divided by the number of spin updates was 28 ns on a Cray X-MP.

\section*{Appendix C: Observables with Improved Overlap}
In a `conventional' correlation length measurement one studies the
exponential decay of correlators $<A_t A_{t+\tau}>$, where $A_t$ is an
observable that depends on the configuration of a single time slice $t$.
On an infinitely long lattice, the correlator can be written as
\be\label{cor}
  < A_t A_{t+\tau}>=
  \frac{\langle 0 \vert A  T^{\tau} A \vert 0 \rangle}
  {\langle 0 \vert T^{\tau} \vert 0 \rangle} \,  .
\ee
Expanding in terms of eigenfunctions of the
transfer matrix one obtains
\be
 A  \vert 0 \rangle  = \sum_i a_i \vert i \rangle \, ,
\ee
where $ a_i = \langle i \vert A \vert 0 \rangle $.
Inserting this in eq.\ (\ref{cor}), one obtains
\be
  <A_t A_{t+\tau}>=
  \frac{\sum_i a_i^2 \lambda_i^{\tau}}{\lambda_0^{\tau}}
  = \sum_i a_i^2 \exp(-m_i \tau) \, ,
\ee
where the masses are defined by
$m_i = -\ln (\frac{\lambda_i}{\lambda_0}) $.
For a general observable $A$ the $a_i$ will be nonzero for
all $i$.

In order to improve this situation one could
think of constructing observables
where only one $a_i$ is non-zero.
First recall that observables correspond to operators which are
diagonal in the basis of configuration states $\vert \phi \rangle$.
This means that
\be
 \langle \phi \vert A \vert \phi' \rangle =
 \langle \phi \vert A \vert \phi \rangle  \, \delta_{\phi,\phi'}
 \equiv  A_{\phi} \, \delta_{\phi,\phi'} \, .
\ee
The condition that $a_i$ should be the only non-vanishing vacuum
overlap is equivalent to $A \vert 0 \rangle = \vert i \rangle$, or
\be
A_{\phi} = \frac{\langle \phi \vert i \rangle}
                {\langle \phi \vert 0 \rangle} \, .
\ee
The problem with this equation is that we do not know
the wave functions $\langle \phi \vert i \rangle$
and $\langle \phi \vert 0 \rangle$ exactly.

However, we know the eigenfunctions $\vert i \RR$ of the effective
transfer matrix. Let us embed them into the Hilbert space of the
full model. We define
\be
\vert i' \rangle \equiv
\sum_M \vert M \rangle \LL M \vert i \RR \, .
\ee
Note that (compare \ section \ref{recover}) this projection leaves
the vacuum invariant, i.e.
$\langle i' \vert 0 \rangle = \LL i \vert 0 \RR $.
As improved observable we now consider the ratio
\be
A_{\phi}' =
\frac{\langle \phi \vert i' \rangle}
     {\langle \phi \vert 0  \rangle} \, .
\ee
One can easily convince oneself that $A_{\phi}'$ can be expressed
in terms of the effective wavefunctions $\vert i \RR$ as
\be
A_{\phi}' = \frac{\LL M(\phi) \vert i \RR}
                 {\LL M(\phi) \vert 0 \RR} \, .
\ee
The relation of our approach with that of \cite{Kronfeld}
 becomes apparent when one identifies: \\
 1) Our $\delta_{\Phi_t,M}$ with  Kronfeld's $\Phi_r^{(i)} (t)$
    in his eq. (4.1),  where the label  $M$ corresponds to $(i)$.
   Note that our  correlation matrix  is already diagonal  for
   distance $0$. \\
 2) Our $ A_{\phi}'$ corresponds to Kronfeld's $z_n^{(i)}$ defined in
    eq.\ (4.6).  \\
 3) The  diagonalization of the
    effective transfer matrix  has its counterpart in the
    variation of  $C_{n,r} (t)$.

\newpage

\listoftables

\section*{Figure Captions}

\vskip0.5cm\noindent
\ul{{\bf Figure 1}}
Overlaps $c_i = \sum_M \langle M \vert i \rangle^2$ for the three
lowest states of the 2-dimensional Ising model with $\Ns = 6$, $8$
and $10$.

\vskip0.5cm\noindent
\ul{{\bf Figure 2}}
Eigenstates of the effective transfer matrix $\Teffl$ for
$l= 5$, $10$ and $20$. $\Ns$ is $64$, and $\beta=\beta_c$.
The full line is the ground-state, the dotted line
is $\vert 0,a \RR$, the dashed line shows $\vert 1,s \RR$,
and the dash-dotted line gives $\vert 1,a \RR$.

\vskip0.5cm\noindent
\ul{{\bf Figure 3}}
Comparison of the convergence of the `effective correlation
lengths' $\xna^{{\rm eff},t}$, `improved' and `standard'.

\vskip0.5cm\noindent
\ul{{\bf Figure 4}}
Comparison of the convergence of the `effective correlation
lengths' $\xes^{{\rm eff},t}$, `improved' and `standard'.

\vskip0.5cm\noindent
\ul{{\bf Figure 5}}
The tunnelling correlation length $\xna$, measured from the $\Ns=64$
lattices.

\vskip0.5cm\noindent
\ul{{\bf Figure 6}}
Eigenstates of the effective transfer matrix $\Teffl$ for
$l= 5$, $10$ and $20$. $\Ns$ is $64$, and $\beta=0.47$.
The full line is the ground-state, the dotted line
is $\vert 0,a \RR$, the dashed line shows $\vert 1,s \RR$,
and the dash-dotted line gives $\vert 1,a \RR$.

\newpage

\begin{table}
\small
\caption[Exact $\xi^{(l)}$ for $\Ns=4$, $\Nt=\infty$ rule]
        {\label{ns4infty}
        as obtained from the exact effective
        transfer matrix for $\Ns=4$,
        using the $\Nt=\infty$ rule}
\begin{center}
\begin{tabular}{|c|c|r|r|r|r|}
\hline
$\beta$ &$l$&
\mc{1}{c|}{$\xna$} &
\mc{1}{c|}{$\xes$} &
\mc{1}{c|}{$\xea$} &
\mc{1}{c|}{$\xzs$} \\
\hline\hline
 0.30 & 1 &   1.50263 & 0.51838 & 0.28378 & 0.21878 \\
 0.30 & 2 &   1.50263 & 0.51880 & 0.28378 & 0.23855 \\
 0.30 & 4 &   1.50263 & 0.51908 & 0.28378 & 0.26000 \\
 0.30 & 8 &   1.50263 & 0.51924 & 0.28379 & 0.28450 \\
 0.30 &4,8&   1.50263 & 0.51940 & 0.28379 & 0.30900 \\
\hline
 0.30 & e &   1.50263 & 0.51940 & 0.28378 & 0.28868  \\
\hline\hline
 0.45 & 1 &   5.41722 & 0.66675 & 0.35470 & 0.27231 \\
 0.45 & 2 &   5.41722 & 0.66726 & 0.35470 & 0.28238 \\
 0.45 & 4 &   5.41722 & 0.66756 & 0.35470 & 0.29141 \\
 0.45 & 8 &   5.41722 & 0.66771 & 0.35470 & 0.29822 \\
 0.45 &4,8&   5.41722 & 0.66786 & 0.35470 & 0.30503 \\
\hline
 0.45 & e &   5.41722 & 0.66786 & 0.35470 & 0.30243 \\
\hline\hline
 0.60 & 1 &  26.11653 & 0.53669 & 0.34962 & 0.28156 \\
 0.60 & 2 &  26.11653 & 0.53705 & 0.34962 & 0.28537 \\
 0.60 & 4 &  26.11653 & 0.53726 & 0.34962 & 0.28802 \\
 0.60 & 8 &  26.11653 & 0.53736 & 0.34962 & 0.28907 \\
 0.60 &4,8&  26.11653 & 0.53746 & 0.34962 & 0.29012 \\
\hline
 0.60 & e &  26.11653 & 0.53747 & 0.34962 & 0.29087 \\
\hline
\end{tabular}
\end{center}
\end{table}
\begin{table}
\small
\caption[Exact $\xi^{(l)}$ for $\Ns=8$, $\Nt=\infty$ rule]
            {\label{ns8infty}
            Correlation lengths $\xi^{(l)}$
            as obtained from the exact effective
            transfer matrix for $\Ns=8$,
            using the $\Nt=\infty$ rule}
\begin{center}
\begin{tabular}{|c|c|r|r|r|r|}
\hline
$\beta$ &$l$&
\mc{1}{c|}{$\xna$} &
\mc{1}{c|}{$\xes$} &
\mc{1}{c|}{$\xea$} &
\mc{1}{c|}{$\xzs$} \\
\hline\hline
 0.30 & 1 &   1.57471 & 0.67038 & 0.38527 & 0.26235 \\
 0.30 & 2 &   1.57479 & 0.67318 & 0.38709 & 0.27564 \\
 0.30 & 4 &   1.57484 & 0.67574 & 0.38861 & 0.31248 \\
 0.30 & 8 &   1.57486 & 0.67740 & 0.38958 & 0.35499 \\
 0.30 &4,8&   1.57488 & 0.67906 & 0.39055 & 0.39750 \\
\hline
 0.30 & e &   1.57488 & 0.67913 & 0.39058 & 0.41348 \\
\hline\hline
 0.45 & 1 &  12.25453 & 1.25922 & 0.62386 & 0.40659 \\
 0.45 & 2 &  12.25990 & 1.26925 & 0.62784 & 0.41963 \\
 0.45 & 4 &  12.26276 & 1.27622 & 0.63047 & 0.43690 \\
 0.45 & 8 &  12.26420 & 1.28000 & 0.63191 & 0.45234 \\
 0.45 &4,8&  12.26564 & 1.28378 & 0.63335 & 0.46796 \\
\hline
 0.45 & e &  12.26565 & 1.28381 & 0.63338 & 0.47114 \\
\hline\hline
 0.60 & 1 & 418.95213 & 0.70382 & 0.52749 & 0.39890 \\
 0.60 & 2 & 419.01927 & 0.71339 & 0.53172 & 0.40984 \\
 0.60 & 4 & 419.05376 & 0.71861 & 0.53417 & 0.41564 \\
 0.60 & 8 & 419.07105 & 0.72131 & 0.53546 & 0.41864 \\
 0.60 &4,8& 419.08834 & 0.72401 & 0.53675 & 0.42164 \\
\hline
 0.60 & e & 419.08835 & 0.72404 & 0.53678 & 0.42169 \\
\hline
\end{tabular}
\end{center}
\end{table}

\begin{table}
\small
\caption[Exact $\xi^{(l)}$ for $\Ns=4$, symmetric periodic lattice rule]
            {\label{ns4periodic}
            Correlation lengths $\xi^{(l)}$
            as obtained from the exact effective
            transfer matrix for $\Ns=4$,
            using the symmetric periodic lattice rule}
\begin{center}
\begin{tabular}{|c|c|r|r|r|r|r|r|r|r|}
\hline
$\beta$ &$l$&
\mc{1}{c|}{$\xna$} &
\mc{1}{c|}{$\xes$} &
\mc{1}{c|}{$\xea$} &
\mc{1}{c|}{$\xzs$} \\
\hline\hline
  0.30 & 1 &   1.50167 &  0.51944 & 0.28375 & 0.22264 \\
  0.30 & 2 &   1.50263 &  0.51880 & 0.28378 & 0.23901 \\
  0.30 & 4 &   1.50263 &  0.51908 & 0.28378 & 0.26007 \\
  0.30 & 8 &   1.50263 &  0.51924 & 0.28378 & 0.28203 \\
  0.30 &4,8&   1.50263 &  0.51940 & 0.28378 & 0.30399 \\
\hline
  0.30 & e &   1.50263 &  0.51940 & 0.28378 & 0.28868  \\
\hline\hline
  0.45 & 1 &   5.41132 &  0.66909 & 0.35467 & 0.27595 \\
  0.45 & 2 &   5.41713 &  0.66737 & 0.35470 & 0.28348 \\
  0.45 & 4 &   5.41722 &  0.66756 & 0.35470 & 0.29205 \\
  0.45 & 8 &   5.41722 &  0.66771 & 0.35470 & 0.29564 \\
  0.45 &4,8&   5.41722 &  0.66786 & 0.35470 & 0.29923 \\
\hline
  0.45 & e &   5.41722 &  0.66786 & 0.35470 & 0.30243 \\
\hline\hline
  0.60 & 1 &  26.10328 &  0.53833 & 0.34962 & 0.28395 \\
  0.60 & 2 &  26.11621 &  0.53726 & 0.34962 & 0.28641 \\
  0.60 & 4 &  26.11653 &  0.53726 & 0.34962 & 0.28834 \\
  0.60 & 8 &  26.11653 &  0.53736 & 0.34962 & 0.29040 \\
  0.60 &4,8&  26.11653 &  0.53746 & 0.34962 & 0.29246 \\
\hline
  0.60 & e &  26.11653 &  0.53747 & 0.34962 & 0.29087 \\
\hline
\end{tabular}
\end{center}
\end{table}
\begin{table}
\small
\caption[Exact $\xi^{(l)}$ for $\Ns=8$, symmetric periodic lattice rule]
         {\label{ns8periodic}
         Correlation lengths $\xi^{(l)}$
         as obtained from the exact effective
         transfer matrix for $\Ns=8$,
         using the symmetric periodic lattice rule}
\begin{center}
\begin{tabular}{|c|c|r|r|r|r|}
\hline
$\beta$ &$l$&
\mc{1}{c|}{$\xna$} &
\mc{1}{c|}{$\xes$} &
\mc{1}{c|}{$\xea$} &
\mc{1}{c|}{$\xzs$} \\
\hline
  0.30 & 1 &   1.56250 & 0.67186 &  0.26386 & 0.38557 \\
  0.30 & 2 &   1.57475 & 0.67310 &  0.27658 & 0.38685 \\
  0.30 & 4 &   1.57484 & 0.67573 &  0.31295 & 0.38815 \\
  0.30 & 8 &   1.57486 & 0.67739 &  0.35092 & 0.38818 \\
  0.30 &4,8&   1.57488 & 0.67905 &  0.38889 & 0.38821 \\
\hline
  0.30 & e &   1.57488 & 0.67913 &  0.39058 & 0.41348 \\
\hline\hline
  0.45 & 1 &  11.68811 & 1.29494 &  0.41267 & 0.63080 \\
  0.45 & 2 &  12.22444 & 1.27367 &  0.42339 & 0.62794 \\
  0.45 & 4 &  12.25978 & 1.27677 &  0.44527 & 0.62700 \\
  0.45 & 8 &  12.26354 & 1.27856 &  0.52313 & 0.65118 \\
  0.45 &4,8&  12.26730 & 1.28035 &  0.60099 & 0.67536 \\
\hline
  0.45 & e &  12.26565 & 1.28381 &  0.63338 & 0.47114 \\
\hline\hline
  0.60 & 1 & 404.80434 & 0.73313 &  0.53976 & 0.40926  \\
  0.60 & 2 & 417.52452 & 0.72285 &  0.53477 & 0.41495  \\
  0.60 & 4 & 418.82046 & 0.72058 &  0.53259 & 0.42259  \\
  0.60 & 8 & 418.96946 & 0.71637 &  0.58641 & 0.53098  \\
  0.60 &4,8& 419.11846 & 0.71216 &  0.64023 & 0.63937  \\
\hline
  0.60 & e & 419.08835 & 0.72404 &  0.53678 & 0.42169  \\
\hline
\end{tabular}
\end{center}
\end{table}

\begin{table}
\small
\caption[$\xi^{(l)}$ at $\beta_c$ from MC for $\Ns=16$, $\Nt=512$]
            {\label{mcl16} Estimates $\xi^{(l)}$ for $\Ns=16$, $\Nt=512$.
            The effective transfer matrix was determined by Monte Carlo,
            using the $\Nt=\infty$ rule. The coupling is $\beta=\beta_c$}
\begin{center}
\begin{tabular}{|r|l|l|l|l|}
\hline
\mc{1}{|c|}{$l$}    &
\mc{1}{c|}{$\xna$} &
\mc{1}{c|}{$\xes$} &
\mc{1}{c|}{$\xea$} &
\mc{1}{c|}{$\xzs$} \\
\hline
  1 &  20.10(8) & 2.402(4)  & 1.147(2)  & 0.704(1) \\
  2 &  20.17(7) & 2.451(4)  & 1.171(3)  & 0.726(2) \\
  3 &  20.22(7) & 2.480(6)  & 1.181(5)  & 0.756(8) \\
  4 &  20.25(7) & 2.499(7)  & 1.193(7)  & 0.76(1)  \\
  5 &  20.26(7) & 2.509(10) & 1.190(15) &          \\
  6 &  20.27(7) & 2.511(12) &           &          \\
  7 &  20.28(7) & 2.518(15) &           &          \\
  8 &  20.29(8) & 2.515(20) &           &          \\
\hline\hline
  e &  20.339   & 2.555     & 1.213     & 0.8730   \\
\hline\hline
 1,2 & 20.25(7) & 2.502(6)  & 1.195(5)  & 0.749(6) \\
 2,4 & 20.32(8) & 2.55(1)   & 1.215(15) &          \\
 3,6 & 20.32(9) & 2.54(2)   &           &          \\
 4,8 & 20.34(9) &           &           &          \\
\hline
\end{tabular}
\end{center}
\end{table}

\begin{table}
\small
\caption[$\xi^{(l)}$ at $\beta_c$ from MC for $\Ns=32$, $\Nt=1024$]
            {\label{mcl32} Estimates $\xi^{(l)}$ for $\Ns=32$, $\Nt=1024$.
            The effective transfer matrix was determined by Monte Carlo,
            using the $\Nt=\infty$ rule. The coupling is $\beta=\beta_c$}
\begin{center}
\begin{tabular}{|r|l|l|l|l|}
\hline
\mc{1}{|c|}{$l$}    &
\mc{1}{c|}{$\xna$} &
\mc{1}{c|}{$\xes$} &
\mc{1}{c|}{$\xea$} &
\mc{1}{c|}{$\xzs$} \\
\hline
  1 &  39.28(12) & 4.44(1) & 2.099(2) & 1.266(2)  \\
  2 &  39.86(13) & 4.61(1) & 2.176(3) & 1.316(2)  \\
  3 &  40.15(14) & 4.71(1) & 2.225(5) & 1.350(4)  \\
  4 &  40.32(15) & 4.77(1) & 2.259(6) & 1.385(5)  \\
  5 &  40.45(16) & 4.83(1) & 2.283(8) & 1.414(6)  \\
  6 &  40.56(17) & 4.87(1) & 2.31(1)  & 1.445(9)  \\
  8 &  40.68(18) & 4.92(2) & 2.34(2)  &           \\
 10 &  40.75(18) & 4.96(2) & 2.34(3)  &           \\
\hline\hline
  e &  40.727   & 5.097    & 2.404    & 1.7014    \\
\hline\hline
 1,2 &  40.45(15) & 4.78(1) & 2.258(6)  & 1.371(4) \\
 2,4 &  40.79(17) & 4.96(2) & 2.348(11) & 1.461(9) \\
 3,6 &  40.97(20) & 5.05(2) & 2.395(20) & 1.55(2)  \\
 4,8 &  41.05(22) & 5.08(2) & 2.42(4)   &          \\
 5,10 & 41.05(22) & 5.10(3) &           &          \\
\hline
\end{tabular}
\end{center}
\end{table}

\begin{table}
\small
\caption[$\xi^{(l)}$ at $\beta_c$ from MC for $\Ns=64$, $\Nt=2048$]
            {\label{mcl64} Estimates $\xi^{(l)}$ for $\Ns=64$, $\Nt=2048$.
            The effective transfer matrix was determined by Monte Carlo,
            using the $\Nt=\infty$ rule. The coupling is $\beta=\beta_c$}
\begin{center}
\begin{tabular}{|r|l|l|l|l|}
\hline
\mc{1}{|c|}{$l$}    &
\mc{1}{c|}{$\xna$} &
\mc{1}{c|}{$\xes$} &
\mc{1}{c|}{$\xea$} &
\mc{1}{c|}{$\xzs$} \\
\hline
  1 & 73.62(31) &  8.13(1) & 3.783(5) & 2.271(3) \\
  2 & 76.00(31) &  8.54(2) & 3.985(4) & 2.396(5) \\
  3 & 77.32(32) &  8.82(2) & 4.119(6) & 2.483(4) \\
  4 & 78.15(31) &  9.01(2) & 4.215(7) & 2.547(5) \\
  5 & 78.74(33) &  9.15(2) & 4.291(8) & 2.598(7) \\
  6 & 79.17(34) &  9.27(2) & 4.346(9) & 2.641(9) \\
  7 & 79.50(34) &  9.36(2) & 4.39(1)  & 2.671(1) \\
  8 & 79.77(34) &  9.44(2) & 4.43(1)  & 2.70(1)  \\
  9 & 79.98(35) &  9.50(2) & 4.46(1)  & 2.73(1)  \\
 10 & 80.13(36) &  9.56(2) & 4.49(2)  & 2.76(2)  \\
 12 & 80.37(36) &  9.64(2) & 4.54(2)  &          \\
 14 & 80.54(37) &  9.70(3) & 4.57(2)  &          \\
 16 & 80.67(38) &  9.75(3) & 4.61(3)  &          \\
 20 & 80.88(40) &  9.81(4) &          &          \\
\hline\hline
  e & 81.479    & 10.188   & 4.797    &  3.4014  \\
\hline\hline
   1,2 & 78.53(32) &  9.00(2) & 4.210(8) & 2.536(5) \\
   2,4 & 80.43(35) &  9.52(3) & 4.47(1)  & 2.72(1)  \\
   3,6 & 81.12(40) &  9.76(3) & 4.60(2)  & 2.82(2)  \\
   4,8 & 81.46(43) &  9.91(3) & 4.66(2)  & 2.88(2)  \\
  5,10 & 81.57(42) & 10.00(3) & 4.72(3)  & 2.95(3)  \\
  6,12 & 81.59(48) & 10.05(4) & 4.75(3)  &          \\
  7,14 & 81.62(43) & 10.08(4) & 4.77(5)  &          \\
  8,16 & 81.59(43) & 10.09(5) &          &          \\
 10,20 & 81.64(46) & 10.08(7) &          &          \\
\hline
\end{tabular}
\end{center}
\end{table}

\begin{table}
\small
\caption[Conventional correlation length estimates at $\beta_c$]
            {\label{conv} `Conventional' correlation length
            estimates $\xna^{{\rm eff},t}$ and
            $\xes^{{\rm eff},t}$ at $\beta=\beta_c$
            from two-slice correlation functions.
            The lattice is $\Ns=64$ by $\Nt=2048$}
\begin{center}
\begin{tabular}{|c|c|c|c|c|c|c|c|}
\hline
$t$    & 10      & 20      & 40      & 80       & & &     \\
\hline
$\xna^{{\rm eff},t}$
& 78.9(5) & 81.4(6) & 82.0(8) & 83.4(14) & & &     \\
\hline
\hline
$t$  & 1     & 3       & 5       & 9       & 11      & 13      &15 \\
\hline
$\xes^{{\rm eff},t}$
& 6.95(2) & 8.33(2) & 9.05(4) & 9.68(6) & 9.81(7) & 9.95(8) & 9.95(11) \\
\hline
\end{tabular}
\end{center}
\end{table}

\begin{table}
\small
\caption[Statistics of the multimagnetical simulations]
        {\label{statbro} Statistics $stat$ for the multimagnetical
        simulations at $\beta=0.47$, given in units
        of one million lattice sweeps}
\begin{center}
\begin{tabular}{|c||c|c||c|c|c||c|c|c|}
\hline
 $\Ns$  & 16 & 16 & 32 & 32 & 32 & 64 & 64 &  64 \\
\hline
 $\Nt$  & 16 & 32 & 16 & 32 & 64 & 32 & 64 & 128 \\
\hline
 $stat$ & 20 & 40 & 40 & 50 & 40 & 50 & 50 &  60 \\
\hline
\end{tabular}
\end{center}
\end{table}

\begin{table}
\small
\caption[$\xi^{(l)}$ at $\beta=0.47$ from MC for $\Ns=16$, asymmetric rule]
            {\label{bro1} Estimates $\xi^{(l)}$
            for $\beta=0.47$ and $\Ns=16$,
            obtained from the effective
            transfer matrix with the asymmetric rule}
\begin{center}
\begin{tabular}{|c|r|l|l|l|}
\hline
\mc{1}{|c|}{$\Nt$} &
\mc{1}{c|}{$l$}    &
\mc{1}{c|}{$\xna$} &
\mc{1}{c|}{$\xes$} &
\mc{1}{c|}{$\xea$} \\
\hline
  16 &  1   & 76.3(7) & 2.041(5) & 1.131(2) \\
  16 &  2   & 76.5(8) & 2.104(5) & 1.158(2) \\
  16 &  4   & 76.7(8) & 2.163(6) & 1.173(6) \\
  16 &  8   & 76.6(8) & 2.21(1)  &          \\
\hline
  32 &  1   & 78.5(4) & 2.013(4) & 1.126(1) \\
  32 &  2   & 78.8(4) & 2.076(4) & 1.155(2) \\
  32 &  4   & 78.9(5) & 2.136(5) & 1.179(4) \\
  32 &  8   & 79.0(5) & 2.18(1)  &          \\
  32 & 16   & 79.1(5) &          &          \\
\hline\hline
   e & $\infty$ & 78.159 & 2.205 &  1.218   \\
\hline
\end{tabular}
\end{center}
\end{table}

\begin{table}
\small
\caption[$\xi^{(l)}$ at $\beta=0.47$ from MC for $\Ns=32$, asymmetric rule]
            {\label{bro2} Estimates $\xi^{(l)}$
            for $\beta=0.47$ and $\Ns=32$,
            obtained from the effective
            transfer matrix with the asymmetric rule}
\begin{center}
\begin{tabular}{|c|r|l|l|l|}
\hline
\mc{1}{|c|}{$\Nt$} &
\mc{1}{c|}{$l$}    &
\mc{1}{c|}{$\xna$} &
\mc{1}{c|}{$\xes$} &
\mc{1}{c|}{$\xea$} \\
\hline
 16 &  1 &   558(9)   &  3.42(2) & 1.914(6) \\
 16 &  2 &   607(10)  &  3.53(2) & 2.034(6) \\
 16 &  4 &   644(11)  &  3.65(2) & 2.172(8) \\
 16 &  8 &   656(13)  &  3.72(2) & 2.27(1)  \\
\hline
 32 &  1 &    718(9)  &  2.68(1) & 1.779(4) \\
 32 &  2 &    728(10) &  2.84(1) & 1.879(5) \\
 32 &  4 &    730(10) &  3.02(1) & 1.982(5) \\
 32 &  8 &    731(10) &  3.18(1) & 2.03(1)  \\
 32 & 16 &    732(11) &  3.30(2) & 2.15(1)  \\
\hline
 64 &  1 &    730(7)  &  2.65(1) & 1.776(3) \\
 64 &  2 &    740(7)  &  2.80(1) & 1.875(3) \\
 64 &  4 &    746(8)  &  2.98(1) & 1.980(4) \\
 64 &  8 &    751(8)  &  3.13(1) & 2.06(1)  \\
 64 & 16 &    757(9)  &          &          \\
 64 & 32 &    760(10) &          &          \\
\hline\hline
 e  &$\infty$ &    753.48  &  3.311   &    2.198 \\
\hline
\end{tabular}
\end{center}
\end{table}

\begin{table}
\small
\caption[$\xi^{(l)}$ at $\beta=0.47$ from MC for $\Ns=64$, asymmetric rule]
            {\label{bro3} Estimates $\xi^{(l)}$
            for $\beta=0.47$ and $\Ns=64$,
            obtained from the effective
            transfer matrix with the asymmetric rule}
\begin{center}
\begin{tabular}{|r|r|l|l|l|}
\hline
\mc{1}{|c|}{$\Nt$} &
\mc{1}{c|}{$l$}    &
\mc{1}{c|}{$\xna$} &
\mc{1}{c|}{$\xes$} &
\mc{1}{c|}{$\xea$} \\
\hline
  32 &  2  & 25800(900)  & 4.00(4) & 2.52(2) \\
  32 &  4  & 30600(900)  & 4.15(4) & 2.80(2) \\
  32 &  8  & 33800(1100) & 4.29(4) & 3.12(2) \\
  32 & 16  & 35400(1900) & 4.38(4) & 3.35(3) \\
\hline
  64 &  4  & 38700(2100) & 3.15(6) & 2.60(3) \\
  64 &  8  & 38600(2000) & 3.46(5) & 2.84(4) \\
  64 & 16  & 39500(2000) & 3.72(5) &         \\
  64 & 32  & 41000(2200) &         &         \\
\hline
 128 & 16  & 41500(1400) &         &         \\
 128 & 32  & 42800(1500) &         &         \\
 128 & 64  & 43500(1600) &         &         \\
\hline\hline
  e  & $\infty$ & 44014.4     & 4.002   & 3.311   \\
\hline
\end{tabular}
\end{center}
\end{table}

\end{document}